\documentclass[a4paper]{jpconf}%
\usepackage{graphicx}
\usepackage{amsmath, amssymb}
\usepackage{epstopdf}
\usepackage{amsmath}
\usepackage{amsfonts}
\usepackage{amssymb}%
\begin{document}

\title{Three-flavor chiral effective model with four baryonic multiplets within the
mirror assignment}
\author{L Olbrich$^{1}$, M Z\'{e}t\'{e}nyi$^{2}$, and F Giacosa$^{1,3}$ }

\begin{abstract}
We study three-flavor octet baryons by using the so-called extended Linear
Sigma Model (eLSM). Within a quark-diquark picture, the requirement of a
mirror assignment naturally leads to the consideration of four spin-$\frac
{1}{2}$ baryon multiplets. A reduction of the Lagrangian to the two-flavor
case leaves four doublets of nucleonic states which mix to form the
experimentally observed states $N(939)$, $N(1440)$, $N(1535)$ and $N(1650)$.
We determine the parameters of the nucleonic part of the Lagrangian from a fit
to masses and decay properties of the aforementioned states. By tracing their
masses when chiral symmetry is restored, we conclude that the pairs $N(939)$,
$N(1535)$ and $N(1440)$, $N(1650)$ form chiral partners.

\end{abstract}

\address{
	$^1$ Institute for Theoretical Physics, Goethe University, Frankfurt am Main, Germany \\    $^2$ Wigner Research Center for Physics, Budapest, Hungary \\     $^3$ Institute of Physics, Jan Kochanowski University, Kielce, Poland}

\ead{
	olbrich@th.physik.uni-frankfurt.de,
	zetenyi.miklos@wigner.mta.hu,
	fgiacosa@ujk.edu.pl}

\section{Introduction}

\label{sec:introduction} In these proceedings, which are based on the results
of Ref.\;\cite{Olbrich}, we focus our attention on baryons consisting of the
light quarks $u,$ $d,$ and $s$ and with quantum numbers $J^{P}=\frac{1}{2}%
^{+}$ and $J^{P}=\frac{1}{2}^{-}$. First and foremost, this includes the
nucleon $N(939)$ and the resonances $N(1440)$, $N(1535)$, $N(1650)$, but also
$\Lambda$, $\Sigma,$ and $\Xi$ resonances \cite{PDG}. 

The fundamental force of nature describing baryons (and hadrons in general) is
Quantum Chromodynamics (QCD), whose Lagrangian is given in terms of quarks and
gluons by (e.g. Ref.\;\cite{weisebook}):
\begin{equation}
\mathcal{L}_{\text{QCD}}=\bar{q}\bigl(i\gamma_{\mu}D^{\mu}-m\bigr)q-\frac
{1}{2}\Tr(\mathcal{G}_{\mu\nu}\mathcal{G^{\mu\nu}})\text{ ,}\label{lqcd}%
\end{equation}
where $D^{\mu}=\partial_{\mu}+ig\mathcal{A}_{\mu}$ is the covariant derivative
and $g$ is the coupling ``constant'' parametrizing the interaction of the quark 
fields $q$ with a gluon field $\mathcal{A}_{\mu}$. 
The latter is associated with the $SU(3)_{c}$ gauge field
$\mathcal{A}^{\mu}=A_{a}^{\mu}T^{a}$ for $a=1,2,\dots,8$ ], $T^{a}$ being the
$SU(3)$ generators. The Yang-Mills field-strength tensor is $\mathcal{G}%
_{\mu\nu}=i[D_{\mu},D_{\nu}]/g$. Definitely, the QCD Lagrangian is elegant,
compact, and contains only few parameters (the bare quark masses and the
coupling $g$). However, a more detailed study within the framework of
renormalization shows that the coupling ``constant'' of strong interaction depends on momentum scale
and becomes arbitrary large in the low-energy regime. This fact forbids the usual
approach of perturbation theory and causes QCD to be not analytically
solvable. On top of that, confinement implies that only ``white'', i.e.\ color singlet states (the
hadrons) are the asymptotic states of the theory, and only these can be measured in detectors.

In order to describe hadrons, theoretical physicists had to look for alternatives. One such possibility is given by effective approaches to QCD, such as chiral perturbation theory (e.g. Refs.\;\cite{scherer}) or linear Sigma Models (see below). In this process one uses Lagrangians which
contain hadrons as degrees of freedom, instead of quarks and gluons as in QCD.
To get the basic idea of an effective model, we
make a short digression by asking under which conditions a glass of water
freezes. The answer is easy: it depends on the temperature of water.
But why is this answer so simple? Actually, the exact description of the
system requires to solve coupled differential equations for each molecule. This is
indeed very complicated, actually impossible also for supercomputers. However,
by using the effective description provided by thermodynamics, one can solve
(some) problems in a much simpler way. In doing so, we lost the information
about the motion of each molecule, but we retained the information that is
important for describing the whole system.

Let us then turn back to hadrons with half-integer spin: baryons. A baryon is
a very complicated system of quarks and gluons arising from Eq.\;(\ref{lqcd}).
For a fast moving baryon, as for instance a proton, one may use (generalized)
parton distribution functions to analyze its substructure in terms of quarks
and gluons, see e.g.\ Ref.\;\cite{gpd}. However, for a baryon in its rest frame,
this is a very hard task, possibly even harder than following each molecule of
water mentioned above. The use of effective descriptions highly
simplify the task. Typically, one uses the concept of `constituent quarks': an
almost massless bare quark is dressed by clouds of gluons and quarks and
becomes a quasiparticle with an effective mass of about $300$ MeV
\cite{weisebook}. A baryon is then described as a bound state of three
constituent quarks. Within this framework, also the concept of diquark as the
strong correlation of two quarks is important, because often baryons are also
regarded as quark-diquark objects.

Certain types of effective models, such as linear Sigma Models, are based on
Lagrangians which contain from the very beginning only hadrons (mesons and
baryons). Quark and gluon fields do not appear. Yet, these Lagrangians are
constructed in such a way that (some of the) symmetries of QCD of Eq.\;(\ref{lqcd}) are
taken into account at he composite level. In particular, chiral symmetry
\cite{weisebook} and its spontaneous and explicit breaking are at the basis of
such approaches. Recently, the so-called extended Linear Sigma Model (eLSM)
has been developed for both mesons and baryons. It contains scalar, pseudoscalar, vector, and axial-vector mesons. Moreover, it
shows chiral symmetry and its breaking but also
dilatation symmetry and its anomalous breaking (this is the non-perturbative
origin of an energy scale of QCD, the renowned $\Lambda_{QCD}\simeq200$ MeV).

In the meson sector, the three-flavour case $N_{f}=3$ has been investigated in
detail in Ref.\;\cite{denisNf=3,moreeLSM}. Yet, until recently, the baryonic
sector was only studied for $N_{f}=2$ in Ref.\;\cite{gallas}. Here, following
Ref.\;\cite{Olbrich}, we enlarge (Sec.\;\ref{sec:the model and its implications}) the eLSM to $N_{f}=3$ in the baryonic
sector. This is a non-trivial step which naturally leads to the consideration of four baryonic
octets. Next, (Sec.\;\ref{sec:fit and result}) we consider the limiting case in which only baryons
with quarks $u$ and $d$ are considered and discuss the mixing patterns and the
identification of the chiral partner of the nucleon. Finally, in Sec.\;\ref{sec:summary} we
present our conclusions and outlooks.

\section{The eLSM and its implications}

\label{sec:the model and its implications} The mesonic part of the eLSM
Lagrangian containing (pseudo)scalar and (axial-)vector mesons is given in Refs.\;\cite{denisNf=3,moreeLSM}. The inclusion of baryons was performed for
$N_{f}=2$ in Refs.\;\cite{gallas} and investigated at finite density in Ref.\;\cite{tetraquarkGallas}. Recently, the development of the baryonic eLSM to
$N_{f}=3$ was undertaken in Ref.\;\cite{Olbrich}. The basic idea is to
construct baryonic fields in a chiral quark-diquark picture. For
$N_{f}=3$ diquarks transform as antiquarks, thus one may construct baryons
in a similar way as mesons. This assumption results in matrices with
elements of different flavor content, which are related to the octet-baryonic
fields: \\[-0.5cm]%
\begin{equation}
\underbrace{\left(
\begin{array}
[c]{r}%
\left[  d,s\right]  \\
-\left[  u,s\right]  \\
\left[  u,d\right]
\end{array}
\right)  }_{\text{diquark}}\underbrace{\left(  u,d,s\right)  }_{\text{quark}%
}\hat{=}\left(
\begin{array}
[c]{ccc}%
uds & uus & uud\\
dds & uds & udd\\
dss & uss & uds
\end{array}
\right)  \sim\left(
\begin{array}
[c]{ccc}%
\frac{\Lambda}{\sqrt{6}}+\frac{\Sigma^{0}}{\sqrt{2}} & \Sigma^{+} & p\\
\Sigma^{-} & \frac{\Lambda}{\sqrt{6}}-\frac{\Sigma^{0}}{\sqrt{2}} & n\\
\Xi^{-} & \Xi^{0} & -\frac{2\Lambda}{\sqrt{6}}%
\end{array}
\right)  \ .
\label{barmatrix}
\end{equation}
When enlarging the present discussion by taking into account
the chirality of quarks and diquarks (see Ref.\ \cite{diquarksGiacosa}) and requiring chiral invariant mass terms, we naturally obtain four baryonic multiplets, two of which transform in a
standard way under chiral transformations and two in a so-called
``mirror'' way. These four multiplets are represented by four matrices analogous to
Eq.\;(\ref{barmatrix}). Two of these matrices
labeled $N_{1}$ and $N_{2}$ behave under chiral transformations as
\begin{align}
N_{1R} &  \rightarrow~U_{R}N_{1R}U_{R}^{\dagger}~,\qquad N_{1L}\rightarrow
~U_{L}N_{1L}U_{R}^{\dagger},\nonumber\\[-0.1cm]
N_{2R} &  \rightarrow~U_{R}N_{2R}U_{L}^{\dagger}~,\qquad N_{2L}\rightarrow
~U_{L}N_{2L}U_{L}^{\dagger}~,\label{eq:chiralTrafo N}%
\end{align}
where $U_{L}$ and $U_{R}$ are $3\times3$ representation matrices of
$U(3)_{L}$ and $U(3)_{R}$. The remaining two matrices $M_{1}$ and $M_{2}$ show
a chiral transformation from the left that is `mirror-like' compared to the
aforementioned:
\begin{align}
M_{1R} &  \rightarrow~U_{L}M_{1R}U_{R}^{\dagger}~,\qquad M_{1L}\rightarrow
~U_{R}M_{1L}U_{R}^{\dagger}\nonumber\\[-0.1cm]
M_{2R} &  \rightarrow~U_{L}M_{2R}U_{L}^{\dagger}~,\qquad M_{2L}\rightarrow
~U_{R}M_{2L}U_{L}^{\dagger}~.\label{eq:chiralTrafo M}%
\end{align}
These transformations comply a so-called ``mirror
assignment'' \cite{mirror} which allows one to introduce chirally invariant baryon mass terms 
in the Lagrangian. The Lagrangian describing these baryonic fields and their interactions with mesonic degrees of freedom is
invariant under chiral symmetry $U(3)_{R}\times U(3)_{L}$ as well as parity
and charge-conjugation transformations \cite{Olbrich}. It reads
\begin{align}
\mathcal{L}_{N_{f}=3,bar}= &  \quad\Tr\left\{  \bar{N}_{1L}i\gamma_{\mu}%
D_{2L}^{\mu}N_{1L}+\bar{N}_{1R}i\gamma_{\mu}D_{1R}^{\mu}N_{1R}+\bar{N}%
_{2L}i\gamma_{\mu}D_{1L}^{\mu}N_{2L}+\bar{N}_{2R}i\gamma_{\mu}D_{2R}^{\mu
}N_{2R}\right\}  \nonumber\\
&  +\Tr\left\{  \bar{M}_{1L}i\gamma_{\mu}D_{4R}^{\mu}M_{1L}+\bar{M}%
_{1R}i\gamma_{\mu}D_{3L}^{\mu}M_{1R}+\bar{M}_{2L}i\gamma_{\mu}D_{3R}^{\mu
}M_{2L}+\bar{M}_{2R}i\gamma_{\mu}D_{4L}^{\mu}M_{2R}\right\}  \nonumber\\
&  -g_{N}\Tr\left\{  \bar{N}_{1L}\Phi N_{1R}+\bar{N}_{1R}\Phi^{\dagger}%
N_{1L}+\bar{N}_{2L}\Phi N_{2R}+\bar{N}_{2R}\Phi^{\dagger}N_{2L}\right\}
\nonumber\\
&  -g_{M}\Tr\left\{  \bar{M}_{1L}\Phi^{\dagger}M_{1R}+\bar{M}_{1R}\Phi
M_{1L}+\bar{M}_{2L}\Phi^{\dagger}M_{2R}+\bar{M}_{2R}\Phi M_{2L}\right\}
\nonumber\\
&  -m_{0,1}\Tr\bigl\{\bar{N}_{1L}M_{1R}+\bar{M}_{1R}N_{1L}+\bar{N}_{2R}%
M_{2L}+\bar{M}_{2L}N_{2R}\bigr\}\nonumber\\
&  -m_{0,2}\Tr\bigl\{\bar{N}_{1R}M_{1L}+\bar{M}_{1L}N_{1R}+\bar{N}_{2L}%
M_{2R}+\bar{M}_{2R}N_{2L}\bigr\}\nonumber\\
&  -\kappa_{1}\Tr\left\{  \bar{N}_{1R}\Phi^{\dagger}N_{2L}\Phi+\bar{N}%
_{2L}\Phi N_{1R}\Phi^{\dagger}\right\}  -\kappa_{1}^{\prime}\Tr\left\{
\bar{N}_{1L}\Phi N_{2R}\Phi+\bar{N}_{2R}\Phi^{\dagger}N_{1L}\Phi^{\dagger
}\right\}  \nonumber\\
&  -\kappa_{2}\Tr\left\{  \bar{M}_{1R}\Phi M_{2L}\Phi+\bar{M}_{2L}%
\Phi^{\dagger}M_{1R}\Phi^{\dagger}\right\}  -\kappa_{2}^{\prime}\Tr\left\{
\bar{M}_{1L}\Phi^{\dagger}M_{2R}\Phi+\bar{M}_{2R}\Phi M_{1L}\Phi^{\dagger
}\right\}  \text{ .}\label{lag}%
\end{align}
The traces are invariant under cyclic permutation, which
ensures their symmetry under the chiral transformations (\ref{eq:chiralTrafo N})
and (\ref{eq:chiralTrafo M}). The covariant derivatives are given by
$D_{kR}^{\mu}=\partial^{\mu}-ic_{k}R^{\mu}$ and $D_{kL}^{\mu}=\partial^{\mu
}-ic_{k}L^{\mu}$ for $k=1,\ldots,4$, where the left- and right-handed matrices
$L^{\mu}$ and $R^{\mu}$ represent (axial-)vector mesonic degrees of freedom.
(Pseudo)scalar mesonic fields are incorporated via the $\Phi$ matrix. The mesonic matrices transform under
chiral transformations as $R^{\mu}\rightarrow U_{R}R^{\mu}U_{R}^{\dagger}$ ,
$L^{\mu}\rightarrow U_{L}L^{\mu}U_{L}^{\dagger}$ , and $\Phi\rightarrow
U_{L}\Phi U_{R}^{\dagger}$. The mass parameters $m_{0,1}$ and $m_{0,2}$ are particularly important, since they allow to shed
light on the origin of the nucleonic masses. They emerge from
(dilatation-invariant) interactions upon the condensation of glueball and/or a
four-quark states, see e.g.\;Ref.\;\cite{tetraquarkGallas}.

The baryonic fields in Eq.\;(\ref{lag}) are not parity eigenstates, therefore we construct the fields of definite parity,
\begin{equation}
B_{N}=\frac{N_{1}-N_{2}}{\sqrt{2}},\qquad B_{N\ast}=\frac{N_{1}+N_{2}}%
{\sqrt{2}},\qquad B_{M}=\frac{M_{1}-M_{2}}{\sqrt{2}},\qquad B_{M\ast}%
=\frac{M_{1}+M_{2}}{\sqrt{2}}\text{.}\label{eq:def of B fields}%
\end{equation}
where now $B_{N}$ and $B_{M}$ have positive parity and $B_{N\ast}$ and
$B_{M\ast}$ have negative parity. In the limit of zero mixing, $B_{N}$
describes the ground-state baryonic fields of Eq.\;(\ref{barmatrix}),
i.e., $\{N(939)$, $\Lambda(1116),$ $\Sigma(1193),$ $\Xi(1338)\},$ $B_{M}$ the
positive-parity fields $\{N(1440),\Lambda(1600),\Sigma(1660),\Xi(1690)\}$,
$B_{N\ast}$ can be assigned to the negative-parity fields $\{N(1535),$
$\Lambda(1670),\Sigma(1620),\Xi(?)\}$ and, finally, $B_{M\ast}$ to
$\{N(1650),\Lambda(1800),\Sigma(1750),\Xi(?)\}$. 
In general the fields describing physical particles emerge as a mixture of
$B_{N}$, $B_{N\ast}$, $B_{M}$, and $B_{M\ast}$.
The detailed study of this mixing will be performed below for the two-flavor case.

\section{Results}

\label{sec:fit and result} In order to determine the twelve parameters of the model,
we reduce Eq.\;(\ref{lag}) to $N_{f}=2$. This leaves us with four isodoublets
(instead of the baryonic $3\times3$ matrices), which mix to produce the
experimentally observed nucleon $N(939)$, $N(1440)$, $N(1535),$ and $N(1650)$.
For the fit, we use thirteen quantities: masses and decay
widths of the resonances, the axial coupling constant of the nucleon from \cite{PDG}, as well as
the lattice results for the remaining axial coupling constants
\cite{Takahashi}. Using a standard $\chi^{2}$-square procedure we found that
three acceptable and almost equally deep minima exist, see Ref.\ \cite{Olbrich}. The
first two minima lead to small absolute values of $m_{0,1}$ and $m_{0,2}$,
while the third one features absolute values of these constants comparable
with the nucleon's mass (in agreement with the recent study of
Ref.\;\cite{nishihara}).

Quite remarkably, for all three minima the decay ${N(1535)\rightarrow N\eta}$
cannot be described (it is a factor 10 too small \cite{Olbrich}). Thus,
further studies are needed to understand $N(1535)$. It may contain a sizable
admixture of $s\bar{s}$, see Ref.\;\cite{sbars}, or the problem might be connected to the
role of chiral anomaly in the baryonic sector \cite{baryonicAnomaly}. The
assignment of chiral partners can also be investigated by computing the masses
as a function of the chiral condensate $\varphi_{N}$, because masses of chiral
partners become degenerate for $\varphi_{N}\rightarrow0$. For all minima, the result
shows that the masses of the $N(939)$ and $N(1535)$ as well as $N(1440)$ and
$N(1650)$ merge as $\varphi_{N}\rightarrow0$, therefore these form two pairs of chiral partners.

Finally, as an illustration, we present here one of the resulting mixing matrices
(Minimum 1 in \cite{Olbrich}): 
\begin{equation}
\left(
\begin{array}
[c]{c}%
N(939)\\
\gamma^{5}N(1535)\\
N(1440)\\
\gamma^{5}N(1650)%
\end{array}
\right)  =\left(
\begin{array}
[c]{rrrr}%
\mathbf{-0.996} & -0.025 & -0.046 & -0.074\\
0.075 & \mathbf{-0.492} & 0.039 & \mathbf{-0.867}\\
-0.050 & -0.057 & \mathbf{0.995} & 0.073\\
0.010 & \mathbf{0.869} & 0.086 & \mathbf{-0.488}%
\end{array}
\right)  \left(
\begin{array}
[c]{c}%
B_{N}\\
\gamma^{5}B_{N\ast}\\
B_{M}\\
\gamma^{5}B_{M\ast}%
\end{array}
\right)  \;.\label{mixingmatrix}%
\end{equation}
In fact, although the mixing has been determined for $N_{f}=2$ only, our calculations show that it is a good
first approximation for all the members of the octet.

\section{Summary and outlook}

\label{sec:summary} The eLSM is an effective model of QCD whose building
blocks are hadrons. Starting from its Lagrangian, one can perform calculations
that are not possible within QCD. In this work we have generalized the eLSM
to the three-flavor case. Requiring chirally invariant mass terms one is lead to use the so-called ``mirror
assignment'', and naturally obtains four baryonic multiplets. In order to
determine the parameters, we have performed a reduction to the $N_{f}=2$ case
and a fit to experimentally known quantities. Three minima produce
results that are in good agreement with experiment (expect for the decay width
$N(1535)\rightarrow N\eta$). Furthermore, we concluded that the pairs
$N(939)$, $N(1535)$ and $N(1440)$, $N(1650)$ form chiral partners. The most important 
open problem is to decide which of the minima is preferable. To this end, we will
investigate the complete $N_{f}=3$ case by performing an overall fit to
measure physical quantities. As a consequence, interesting information for
both vacuum physics, such as scattering processes involving strange hadrons
\cite{kaon} and at nonzero density, such as the role hyperon in compact stars
\cite{astro}, will be obtained.

\bigskip

\ack L.O.\ acknowledges support by HGS-HIRe/HQM. M.Z. was supported by the
Hungarian OTKA Fund No. K109462 and HIC for FAIR.

\end{document}